\documentclass[aps,prl,twocolumn,superscriptaddress]{revtex4}
\usepackage[utf8]{inputenc}
\usepackage{amsmath}
\usepackage{amsfonts}
\usepackage{amssymb}
\usepackage{color}
\usepackage{graphicx}
\usepackage[normalem]{ulem}
\usepackage{multirow}
\usepackage{tabularx} 
\usepackage{mathrsfs}
\newcolumntype{L}[1]{>{\raggedright\arraybackslash}p{#1}}
\newcolumntype{C}[1]{>{\centering\arraybackslash}p{#1}}
\newcolumntype{R}[1]{>{\raggedleft\arraybackslash}p{#1}}
\def\vk{{\bf k}}

\begin{document}
\title{Topological superconductor in quasi-one-dimensional Tl$_{2-x}$Mo$_6$Se$_6$}

\author{Shin-Ming Huang}
\affiliation {Department of Physics, National Sun Yat-sen University, Kaohsiung 80424, Taiwan}
\author{Chuang-Han Hsu}
\affiliation {Centre for Advanced 2D Materials and Graphene Research Centre, National University of Singapore, 6 Science Drive 2, Singapore 117546}
\affiliation {Department of Physics, National University of Singapore, 2 Science Drive 3, Singapore 117542}
\author{Su-Yang~Xu}
\affiliation {Department of Physics, Massachusetts Institute of Technology, Cambridge, Massachusetts 02139, USA} 
\author{Chi-Cheng~Lee}
\affiliation{Institute for Solid State Physics, The University of Tokyo, Kashiwa 277-8581, Japan}
\author{Shiue-Yuan Shiau}
\affiliation {Centre for Advanced 2D Materials and Graphene Research Centre, National University of Singapore, 6 Science Drive 2, Singapore 117546}
\affiliation {Department of Physics, National University of Singapore, 2 Science Drive 3, Singapore 117542}
\author{Hsin~Lin}
\affiliation {Centre for Advanced 2D Materials and Graphene Research Centre, National University of Singapore, 6 Science Drive 2, Singapore 117546}
\affiliation {Department of Physics, National University of Singapore, 2 Science Drive 3, Singapore 117542}
\affiliation {Institute of Physics, Academia Sinica, Nankang Taipei 11529, Taiwan}
\author{Arun Bansil}
\affiliation {Department of Physics, Northeastern University, Boston, Massachusetts 02115, USA}

\begin{abstract}
We propose that the quasi-one-dimensional molybdenum selenide compound Tl$_{2-x}$Mo$_{6}$Se$_{6}$ is a time-reversal-invariant topological superconductor induced by inter-sublattice pairing, even in the absence of spin-orbit coupling (SOC). No noticeable change in superconductivity is observed in Tl-deficient ($0\leq x \leq 0.1$) compounds. At weak SOC, the superconductor prefers the triplet $d$ vector lying perpendicular to the chain direction and two-dimensional $E_{2u}$ symmetry, which is driven to a nematic order by spontaneous rotation symmetry breaking. The locking energy of the $d$ vector is estimated to be weak and hence the proof of its direction would rely on tunnelling or phase-sensitive measurements. 
\end{abstract}

\date{\today}
\maketitle
\emph{Introduction.}---Topological superconductivity stands out among all the topological phases in part because the Majorana fermions it allows at boundaries are not only fundamentally fascinating but also have potential applications in quantum computation \cite{Read2000,Ivanov2001,Fu2008,Alicea2012,Sato2016}. A crucial element for such superconductors is spin-triplet pairing, or odd-parity pairing in the presence of inversion symmetry \cite{Sato2010,Fu2010}. Prime examples are unconventional superconductors Sr$_2$RuO$_4$ \cite{Maeno2012,Haverkort2008} and UPt$_3$ \cite{Saul1994} and both of them act as chiral superconductors \cite{Xia2006,Strand2010,Schemm2014}. However, their complex or nodal Fermi surface properties make it difficult to host distinct Majorana modes \cite{Maeno2012,Hassinger2017,Saul1994,Strand2010}. One could seek to exploit the proximity effect between the surface states of an topological insulator and an $s$-wave superconductor \cite{Fu2008,Oreg2010,Lutchyn2010,Nadj-Perge2014}. Such a approach requires strong spin-orbit coupling (SOC)  to break the inversion symmetry in order to turn singlet pairing into triplet pairing. Another rather unexpected avenue toward realizing topological superconductors is through doping topological insulators. The electron-doped topological insulator Cu$_x$Bi$_2$Se$_3$ has just been verified as a spin-triplet  superconductor with a critical temperature $T_c\sim3.2$ K \cite{Hor2010,Fu2010,Matano2016}, and possibly carrying a nematic order \cite{Fu2014,Matano2016,Yonezawa2016}.

In this Letter, we propose the non-symmorphic semimetal compound  Tl$_{2-x}$Mo$_{6}$Se$_{6}$ as a spin-triplet topological superconductor with time-reversal symmetry.  Being  a representative of molybdenum selenides $\mathcal{M}_{2}$Mo$_{6}$Se$_{6}$ ($\mathcal{M}$=Na, Rb, In, or Tl), this  compound becomes superconducting with  $T_c \sim 3-6.5$ K \cite{Armici1979,Tarascon1984,Brusetti1988,Petrovic2010}, and is fully gapped according to differential resistance studies \cite{Bergk2011}. Tl deficiency $x$ falls between 0 to 0.1.  Another superconductor in the family is $\mathcal{M}$=In compound with $T_c\sim 2.9$ K \cite{Petrovic2010}. Tl$_{2-x}$Mo$_{6}$Se$_{6}$ superconductor is particularly interesting because a small SOC will suffice to fix the triplet $d$ vector \cite{Mackenzie2000} and yield a two-component order parameter in the $E_{2u}$ irreducible representation. Moreover, the superconductivity is rather insensitive to doping within the available experimental doping range ($0\leq x \leq 0.1$). Similar to Cu$_x$Bi$_2$Se$_3$ superconductor, we expect a concurrent nematic order \cite{Fu2014}, and a nematic vortex from the crossing of nematic domain walls, around which (pseudo)spin-up and spin-down order parameters gain phases of $\pm 2\pi$, will bind the Kramers pair of Majorana modes robust against disorder \cite{Wu2017}.

\begin{figure}[tbp]
\includegraphics[width=0.48\textwidth]{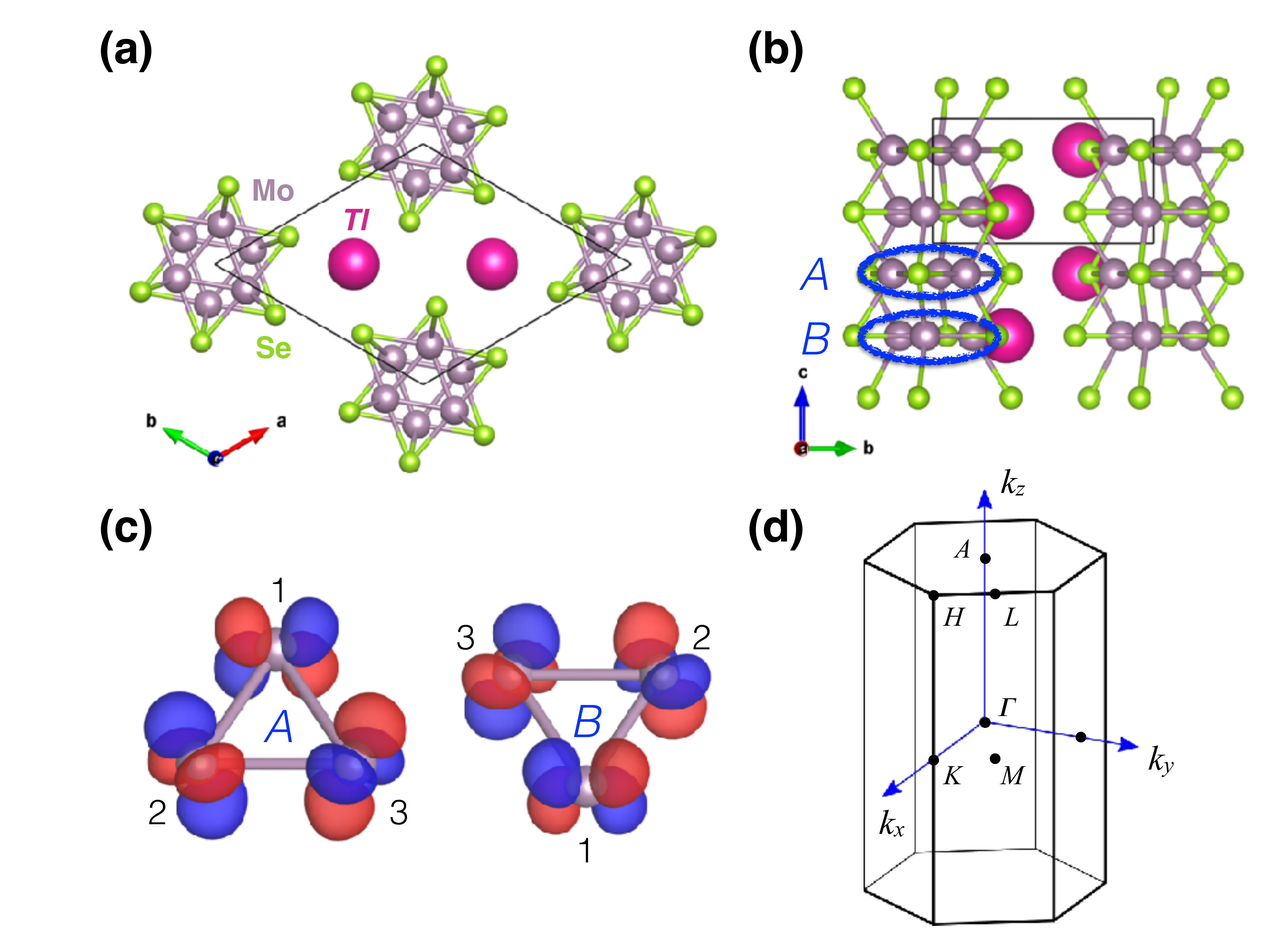} \newline
\caption{(Color online) Hexagonal lattice and Brillouin zone of Tl$_2$Mo$_{6}$Se%
$_{6}$. (a,b) Top/Side view of a unit cell. Large
pink, median purple and small green spheres denote Tl, Mo and Se atoms,
respectively. In a unit cell, two Mo triangles, $A$ and $B$, form the (electronic)
basis. (c) Wannier functions for $A$ and $B$ sublattices come from $d_{xz}$ orbitals and are $C_{3}$-invariant. Red and blue colors in the Wannier function stand for opposite signs. (d) The first Brillouin
zone and high-symmetry $k$ points. $\Gamma$, $M$, $A$, and $L$ are time-reversal invariant momenta.}
\label{fig1}
\end{figure}

\emph{Electronic structure.}---Non-symmorphic compound Tl$_{2-x}$Mo$_{6}$Se$_{6}$ has a hexagonal lattice with
inversion symmetry, characterized by the space group $P6_{3}/m$ (No. 176).  Having highly
anisotropic lattice constants ($a$=8.934 $\mathring{A}$, $c$%
=4.494 $\mathring{A}$), its crystal structure is  quasi-one-dimensional (quasi-1D), with Mo$_{3}$Se$_{3}$ chains arranged in a triangular lattice, as shown in Figs. \ref{fig1}(a,b). A Tl atom is centered in each triangle to couple the Mo$_{3}$Se$_{3}$ chains. It also act as an electron donor, stable in  a $+1$ valence state, much like an alkali atom~\cite%
{Petrovic2010}.  Two inverse closely-packed Mo$_3$ triangles at $3\vec{c}/4$
and $\vec{c}/4$, dubbed $A$ and $B$, form a basis of the Mo$_{3}$Se$_{3}$ chain. Owing to neighboring three Se anions
and one Tl cation, three Mo atoms of a
triangle have to share five $4d$ valence electrons, which implies a half-filled
band in the absence of Tl deficiency. 

\begin{figure}[tbp]
\includegraphics[width=0.4\textwidth]{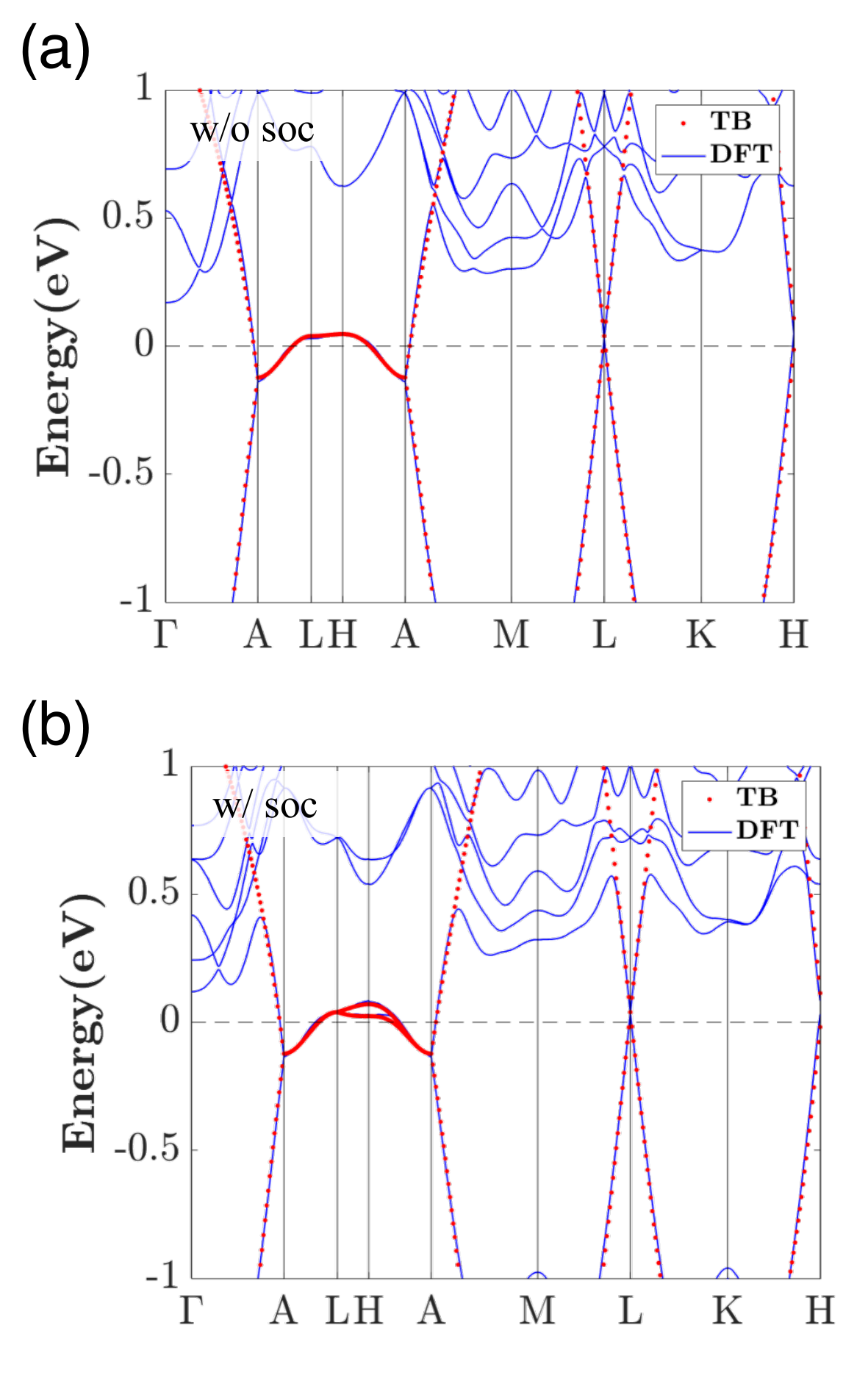} \newline
\caption{(Color online) Band structures of  undoped Tl$_2$Mo$_6$Se$_6$ with (a) or without (b) SOC. Blue lines represent results from first-principles calculations and red dots from fitting the tight-binding model. The Fermi energy is set to zero.}
\label{fig2}
\end{figure}

We have performed first-principles band calculations \cite{wien2k} within density functional theory using generalized
gradient approximation \cite{PBE96}. 
Tl$_{2-x}$Mo$_{6}$Se$_{6}$ band structures are shown in Fig.~\ref{fig2}.
The bands close to the Fermi level mainly come from the Mo $d_{xz}$ orbitals with lobes of their Wannier wave functions pointing toward nearby Se atoms [see Fig. \ref{fig1}(c)]. They are the basis states from which we construct the low-energy Hamiltonians. Two $A$ and $B$ sublattice states can interchange under inversion $\mathcal{I}$, the
center of inversion being in the middle of the two sublattices. They can also
interchange by a two-fold screw operation $\mathcal{S}_{2}$ along the $z$ direction, plus a translation by $\vec{c}/2$. 


Consider first a simple 1D Mo$_{3}$Se$_{3}$ chain in the $z$ direction. (Throughout the paper we shall set lattice constants to unity.)  The two sublattices in a unit cell form symmetric and anti-symmetric states and they modulate with phase $e^{ik_{z}z}$ to be a bonding and a anti-bonding band. At $k_{z}=\pi $, the Bloch-state modulating phase is equal to $(-1)^{z}$ and the bonding and anti-bonding states become identical under inversion, $\mathcal{I}\Psi_{\mathrm{Bond}} = \Psi_{%
\mathrm{Antibond}}$ (up to a phase), so that two bands touch. 

By extending to three dimensions through introducing inter-chain
coupling, the band crossing evolves into a two-dimensional (2D) nodal surface at $k_{z}=\pi $ \cite{Liang2016}. The two-band Hamiltonian reads 
\begin{equation}
\mathcal{H}_{t}(\mathbf{k}) = \varepsilon_0(\mathbf{k}) \sigma_0 + \varepsilon_{1}(\mathbf{k}) \sigma_{1} + \varepsilon_{2}(\mathbf{k}) \sigma_{2},
\end{equation}
where $\sigma_{0}$ and $\sigma_{i}$ are the identity and Pauli matrices for sublattices.
The $\sigma_{3}$ term is forbidden by the $\mathcal{TI}$ symmetry \cite{SM}. Furthermore, $(\mathcal{T S}_{2})^2$ is a unit
lattice translation by $\vec{c}$ and gives $e^{-i k_z}$ by acting
on a Bloch eigenstate. Particularly, $(\mathcal{T S}_{2})^2=-1$ at $k_z=\pi$%
, leading to a double degeneracy for all states, analogous to the
Kramers degeneracy \cite{Liang2016}. As a result, $\varepsilon_1(\mathbf{k})=\varepsilon_2(\mathbf{k})=0$ at $k_z=\pi $. As shown in Fig.~\ref{fig2}(a), states along the A-L-H-A path on $k_z=\pi$ plane are degenerate.

Under time-reversal and inversion symmetries,  the Hamiltonian that includes SOC reads \cite{SM}
\begin{equation} \label{HSOC}
\mathcal{H}(\mathbf{k}) = s_0 \mathcal{H}_{t}(\mathbf{k}) +  \vec{\zeta}(\mathbf{k})\cdot \vec{s}~\sigma_{3},
\end{equation}
where $s_{0}$ and $\vec{s}=(s_1,s_2,s_3)$ are the identity and Pauli matrices for spin. As a consequence of $\mathcal{T}$ and $\mathcal{I}$ symmetries, $\zeta$'s must be odd in $\vk$. This $\sigma_{3}$ term will gap out the surface node. According to  first-principles calculations, the SOC is quite weak for Tl atom, especially along $A-L$, but gets larger for heavier $\mathcal{M}$ atom [see Fig. \ref{fig2}(b)]. Since SOC vanishes at time-reversal invariant momenta, residual point-like band crossings appear at $A$ and $L$ points, namely three-dimensional (3D) Dirac nodes. In particular, the band crossing at $A$ point is a cubic Dirac fermion due to its six-fold symmetry \cite{Liu2017}.

\emph{Theory of superconductivity.}---Under time-reversal symmetry, the Tl$_2$Mo$_{6}$Se$_{6}$ superconducting states, described by the mean-field Hamiltonian
\begin{equation} 
\mathcal{H}_{\mathrm{SC}}(\mathbf{k}) =\left(\begin{smallmatrix}
\mathcal{H}(\mathbf{k}) - \mu & \Delta(\mathbf{k})\\ \Delta(\mathbf{k}) & -[\mathcal{H}(\mathbf{k}) - \mu]\end{smallmatrix}\right) 
\end{equation}
in the basis $\Psi_{\mathbf{k}} = (\psi_{\mathbf{k}\uparrow}^{\mathrm{T}},\psi_{\mathbf{k}\downarrow}^{\mathrm{T}},\psi_{-\mathbf{k}\downarrow}^{\dagger},-\psi_{-\mathbf{k}\uparrow}^{\dagger})^{\mathrm{T}}$, can be classified into even (spin singlet) and odd parity (spin triplet) pairing by
$\mathcal{I}\Delta (\mathbf{k})\mathcal{I}^{-1}=\pm \Delta (-\mathbf{k})$.
For a DIII topological superconductor, the parity of the superconducting gap has to be odd \cite{Sato2010}. In conventional superconductivity, spin singlet/triplet
imposes that the gap function is even/odd in $\vk$. However, additional sublattice
degrees of freedom plays a role much like in the orbital
pairing theory in Fe-based superconductors \cite{Dai2008}: For spin singlet, to make parity even, we can have sublattice and $\vk$ parts to be both even or both odd. For spin triplet, to make parity odd,  we can have sublattice even and $\vk$ odd, or sublattice odd and $\vk$ even. Since $\mathcal{I}$ is a combination of the two-fold screw operation $\mathcal{S}_{2}$  and mirror  operation $\mathcal{M}_z$ (i.e., $z\rightarrow -z$), we can  further classify the superconducting states in each parity  into screw even ($+$) and screw odd ($-$), according to 
\begin{equation*}
\mathcal{S}_{2}(k_{z})\Delta (k_{x},k_{y},k_{z})\mathcal{S}_{2}(k_{z})^{\mathrm{T}}=\pm \Delta (-k_{x},-k_{y},k_{z}).
\end{equation*}%
Similarly for $\mathcal{M}_z$.The full classification of the gap functions according to $C_{6h}$ group is shown in Table \ref{table1}. 

\begin{table*}[tbp]
\begin{center}
\begin{tabular}{|C{1.5cm}|C{1.5cm}C{1.5cm}llC{3cm}|}
\firsthline
Rep. & $\mathcal{S}_2$ & $\mathcal{M}_z$ & Basis functions & $\Delta (\mathbf{k})$ & Susceptibilities\\
\hline \hline
$A_g$ ($\Delta_{1}$) & $+$ & $+$ & $1$ & $ s_0 \otimes \sigma _{0}; \cos \frac{k_z}{2} s_0 \otimes \sigma _{1} $ & $\chi_0,\chi_{01},\chi_{1}$\\
$B_g$ ($\Delta_{2}$) & $-$ & $-$ & $\mathrm{Re} \, k_{+}^{3} k_z ; \mathrm{Im} \, k_{+}^{3} k_z $ & $ s_0 \otimes \sin \frac{k_z}{2} \sigma _{2}$ & $\chi_{2}$\\
$E_{1g}$ & $-2$ & $-2$ & $\left\{ \begin{array}{c} k_{x} k_z\\ k_{y} k_z\end{array}\right\}$ & N/A & \\
$E_{2g}$ & $+2$ & $+2$ & $\left\{ \begin{array}{c} k_{x}^2-k_{y}^2\\ 2k_{x}k_{y}\end{array}\right\}$ & N/A & \\ 
\hline
$A_u$ ($\Delta_{3}$) & $+$ & $-$ & $k_z \hat{z}; k_x \hat{x}+k_y \hat{y};k_x\hat{y}-k_y\hat{x}$ & $s_3 \otimes \sin \frac{k_z}{2} \sigma _{1}$ & $\chi_{3}$ \\
$B_u$ ($\Delta_{4}$) & $-$ & $+$ & $\mathrm{Re} \, k_{+}^3 \hat{z}; \mathrm{Im} \, k_{+}^3 \hat{z}$ & $s_3 \otimes  \cos \frac{k_z}{2} \sigma _{2}$ & $\chi_{4}$\\
$E_{1u}$ ($\Delta_{5}$) & $-2$ & $+2$ & $\left\{ \begin{array}{c} k_{x}\hat{z}\\ k_{y}\hat{z}\end{array}\right\}$;$\left\{ \begin{array}{c} k_{z}\hat{x}\\ k_{z}\hat{y}\end{array}\right\}$ & $\left\{ \begin{array}{c} s_1 \\ s_2 \end{array}\right\} \otimes \cos \frac{k_z}{2} \sigma _{2}$ & $\chi_{5}$ \\
$E_{2u}$ ($\Delta_{6}$) & $+2$ & $-2$ & $\left\{ \begin{array}{c} (k_{x}^2-k_{y}^2) k_z \hat{z}\\ 2k_x k_y k_z \hat{z} \end{array}\right\}$;
$\left\{ \begin{array}{c} k_x\hat{x}- k_y \hat{y} \\ k_{y}\hat{x} + k_{x} \hat{y}\end{array}\right\}$ & $ \left\{ \begin{array}{c} s_1 \\ s_2 \end{array}\right\} \otimes \sin \frac{k_z}{2} \sigma _{1}$ & $\chi_{6}$\\ 
\hline
\end{tabular}
\end{center}
\caption{Classification of gap functions for the interaction in Eq.~(\ref{H_int}) in $C_{6h}$ group. $g$ ($u$) subscript denotes even (odd) parity. $A$'s and $B$'s are 1D representations, $E$'s are 2D representations. The second and third columns display traces of eigenvalues of $\mathcal{S}_2$ and $\mathcal{M}_z$ for every representation. The gap function is
defined by $\Delta (\mathbf{k})= s _{i}\protect\otimes\sum_{\protect j %
=0}^{3}\Delta _{\protect j }(\mathbf{k})\protect\sigma _{\protect j }$. $s_{0}$ stands for spin singlet ($\uparrow\downarrow-\downarrow\uparrow$), and $%
\protect s_{1,2,3}$ stands for spin triplet ($\uparrow\uparrow-\downarrow\downarrow$, $\uparrow\uparrow+\downarrow\downarrow$, $\uparrow\downarrow+\downarrow\uparrow$), respectively. $\sigma_{0}$ and $\sigma_{3}$ ($\sigma_{1}$ and $\sigma_{2}$) terms are intra- (inter-)sublattice pairings. $\protect\sigma _{0}$ and $\protect\sigma %
_{1}$ ($\protect\sigma _{2}$ and $\protect\sigma _{3}$) terms are sublattice even (odd). The basis functions allude to gap functions from the band particles (Here $k$'s are expanded around the $A$ point and $k_{+}=k_x+ik_y$). }
\label{table1}
\end{table*}

We consider attractive, intra-
and inter-sublattice interactions for pairing,  
\begin{align}  \label{H_int}
H_{\mathrm{int}} &= - \sum_{\mathbf{r}} \biggl\{ U \big[ n_{A \uparrow}(%
\mathbf{r})n_{A \downarrow}(\mathbf{r}) + n_{B \uparrow}(\mathbf{r})n_{B
\downarrow}(\mathbf{r}) \big] \\
&+ V \big[ n_{A}(\mathbf{r})n_{B}(\mathbf{r}+\vec{c}/2) + n_{A}(\mathbf{r}%
)n_{B}(\mathbf{r}-\vec{c}/2)\big] \biggr\},  \notag
\end{align}
where $\mathbf{r}$ runs over the Bravais lattice and $n_{X}(\mathbf{r}) =
n_{X \uparrow}(\mathbf{r})+n_{X \downarrow}(\mathbf{r}) $ is the electron
density for sublattice $X$. Interactions between chains are neglected since we expect pairing to be the strongest within a chain. Table \ref{table1} lists six possible pairing symmetries. While the intra-sublattice interactions participate in the $A_{g}$ pairing, the inter-sublattice interactions contribute to all six pairings \cite{SM}. 
Their critical temperatures $T_{c}$ are determined by 
\begin{align}  \label{Tcs}
\begin{split}
 & \det \left[ \left( 
\begin{array}{cc}
\frac{U}{4}\chi_{0}(T_{c}) & \frac{U}{4}\chi_{01}(T_{c}) \\ 
\frac{V}{2}\chi_{01}(T_{c}) & \frac{V}{2}\chi_{1}(T_{c})%
\end{array}
\right)-{\rm I}\right] =0,~~~{\rm for}~~~  \Delta_{1} \\
 &\frac{V}{2}\chi_{i}(T_{c}) =1,~~~{\rm for}~~~ \Delta_{i=2,3,4,5,6}.  
\end{split}%
\end{align}
The pair susceptibility $\chi_i$ is given by 
\begin{equation} \label{susceptibility}
\chi_{i} = \frac{T}{N} \sum_{\mathbf{k},i\omega_{n}} \mathrm{Tr} \big( \Gamma_i(\mathbf{k}) G(\mathbf{k},i \omega_n) \Gamma_i(\mathbf{k})  G(\mathbf{k},-i \omega_n) \big), 
\end{equation}
where $G(\mathbf{k},i \omega_n)=[i\omega_n-\mathcal{H}(\mathbf{k})+\mu]^{-1}$ with $\omega_n=(2n+1)\pi k_BT$ is the fermion Green's function, and  $N$ is the number of lattice sites. The chemical potential $\mu$ is used to simulate doping effect. The vertex functions are $\Gamma_{0}(\mathbf{k}) = s_{0} \sigma_{0}$, $\Gamma_{1}(\mathbf{k}) = s_{0} \sigma_{1} \cos(k_z/2)$, $\Gamma_{2}(\mathbf{k}) = s_{0} \sigma_{2} \sin(k_z/2)$, $\Gamma_{3}(\mathbf{k}) = s_{3} \sigma_{1} \sin(k_z/2)$, $\Gamma_{4}(\mathbf{k}) = s_{3} \sigma_{2} \cos(k_z/2)$, $\Gamma_{5}(\mathbf{k}) = s_{1} \sigma_{2} \cos(k_z/2)$, and $\Gamma_{6}(\mathbf{k}) = s_{1} \sigma_{1} \sin(k_z/2)$. $\chi_{01}$ is obtained by replacing the first $\Gamma_i(\mathbf{k})$ in Eq.~(\ref{susceptibility}) by $\Gamma_{0}(\mathbf{k})$ and the second by $\Gamma_{1}(\mathbf{k})$.

\begin{figure}[tbp]
\includegraphics[width=0.5\textwidth]{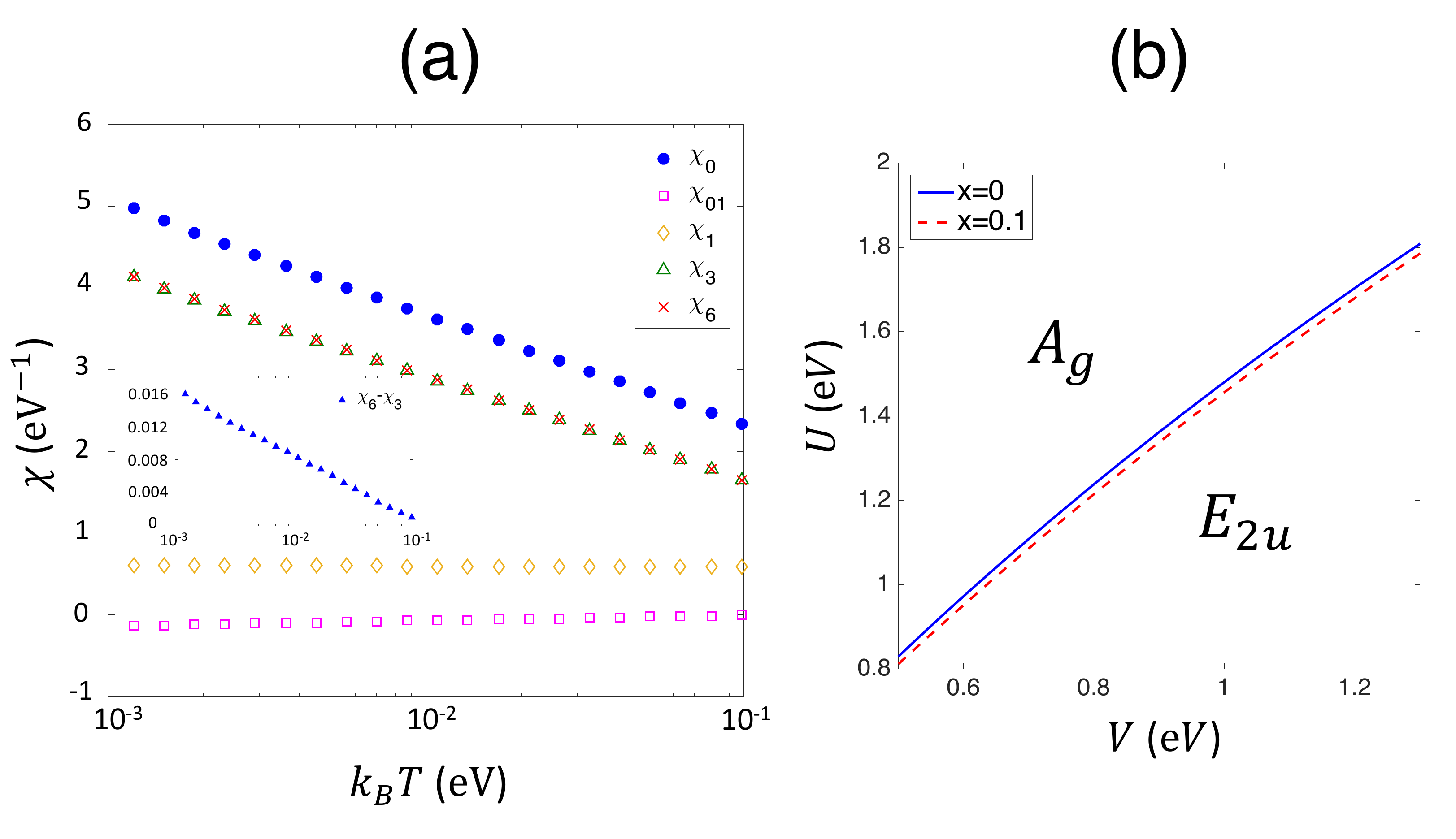} \newline
\caption{(Color online) (a) Pair susceptibilities of Tl$_{2-x}$Mo$_6$Se$_6$ with $x=0$ for relevant channels. Large $\chi_0$, $\chi_3$, and $\chi_6$ correspond to $A_{g}$, $A_{u}$, and $E_{2u}$ states, respectively. $\chi_3$ and $\chi_6$ curves almost overlap. (b) Superconducting phase diagram between $A_{g}$ and $E_{2u}$ states as functions of intra- and inter-sublattice interaction strengths, $U$ and $V$. Doping has weak effect on the phase boundary. The range of $V$ corresponds to $T_c$ from about 10 K to 1000 K (nonlinear relation).}
\label{fig3}
\end{figure}

Adopting the fitting parameters for the band structure in Fig.~\ref{fig2}(b), we computed the pair susceptibility for these six channels [see Fig.~\ref{fig3}(a)]. Three dominant channels are $\chi_0$, $\chi_3$ and $\chi_6$ for,  respectively, 1D $A_g$ and $A_u$, and 2D $E_{2u}$ irreducible representations, all having sublattice-even and screw-even pairings. In Table \ref{table1}, those  for gapful superconductivity show the conventional logarithmic behavior, $\chi\sim \mathcal{N}_{\mathrm{eff}} \ln \left( \Lambda /k_{\mathrm{B}}T\right)$, where $\mathcal{N}_{\mathrm{eff}}$ stands for the effective density of states at the Fermi energy and $\Lambda$ is the energy cutoff. The logarithmic law taken into Eq.~(\ref{Tcs}) determines the critical temperature for superconductivity. 

We first consider $A_g$ and $E_{2u}$ states. Depending on the  interaction strength  ratio $U/V$, the superconductor can fall into $A_g$ or $E_{2u}$ state. From Eq.~(\ref{Tcs}), the condition for $E_{2u}$ to dominate over $A_{g}$ is 
\begin{equation}
\frac{U}{2V} < \frac{ \tilde{\chi}_{1} - 1 }{ \tilde{\chi}_{0} \left( \tilde{\chi}_{1} - 1\right) -\tilde{\chi}_{01}^{2}}
\end{equation}
where $\tilde{\chi}_{0,01,1}=\chi_{0,01,1}(T_{c})/\chi_{6}(T_{c})$; otherwise, $A_{g}$ dominates over $E_{2u}$. The phase diagram is shown in Fig. \ref{fig3}(b). Doping would only slightly  shift the phase boundary and  reduce $T_c$. At present little is known about the origin of pairing and the values of $U$ and $V$. If the attractive interactions are phonon-mediated, the spin triplet state could be stabilized by a weak electronic correlation, as proposed for Cu$_x$Bi$_2$Se$_3$ \cite{Brydon2014}. Then, considering that strong on-site repulsion is common in $4d$ transition metals, the $E_{2u}$ state is likely the winner.

 The critical temperatures for $A_u$ and $E_{2u}$ states  are very close since their $\chi$ difference is small \cite{SM}. In the absence of SOC, the $E_{2u}$ state will be triply degenerate with the $A_u$ state because of SU(2) spin symmetry. The SOC lifts their degeneracy, favoring $E_{2u}$. The solution to the fact is that the highest $T_{c}$ is obtained when the triplet $d$ vector aligns with the SOC field $\vec{\zeta}(\mathbf{k})$ in Eq.~(\ref{HSOC}) \cite{Frigeri2004}. For the Fermi surface located around $k_z=\pi$, $\vec{\zeta}(\mathbf{k})$ and hence the $d$ vector, in principle lies on the $x$-$y$ plane \cite{SM}.

\emph{Nematic order.}---
To study spontaneous  time-reversal and rotation symmetry breaking, we consider the phenomenological Ginzburg-Landau free energy for the spin-triplet $E_{2u}$ state, which reads
\begin{align} \label{GL}
\begin{split}
F=&\alpha \left( |\Psi_+|^2 + |\Psi_-|^2 \right) + \beta_1 \left( |\Psi_+|^2 + |\Psi_-|^2 \right)^2 \\ 
  &+ \beta_2 \left\vert \Psi_{+} \right\vert^2  \left\vert \Psi_{-} \right\vert^2,
\end{split}
\end{align}
where  $\Psi_{\pm} = \Psi_{1} \pm i \Psi_{2}$. The two order parameters for $E_{2u}$, $\Psi_{1}$ and $\Psi_{2}$, correspond to the pairing states $i \langle c^{\dagger}_{A \uparrow} c^{\dagger}_{B \uparrow} - c^{\dagger}_{A \downarrow} c^{\dagger}_{B \downarrow}\rangle$ and $\langle c^{\dagger}_{A \uparrow} c^{\dagger}_{B \uparrow} + c^{\dagger}_{A \downarrow} c^{\dagger}_{B \downarrow}\rangle$, respectively. Below $T_c$ ($\alpha<0$), we obtain $\Psi_{\pm} \neq 0$ and superconductivity occurs. The sign of $\beta_2$ is crucial for determining whether time-reversal symmetry ($\beta_2>0$) or  rotation symmetry ($\beta_2<0$) breaks  \cite{Fu2014}.  We shall rule out the $\mathcal{T}$-breaking scenario since no magnetic moment is found in Tl$_{2-x}$Mo$_6$Se$_6$ \cite{Brusetti1988}. In order to determine the nematic angle $\theta$ in the crystal, we need the sixth-order term in the free energy, given by
\begin{equation}
\delta F_{6} = -(\gamma_{1}+ i \gamma_{2} ) \left(\Psi_{+}^{*}\Psi_{-}\right)^3 + \mathrm{H.c.} 
\end{equation}
where $(\gamma_1,\gamma_2)$ depend on  microscopic models. This term is proportional to $-\sqrt{\gamma_{1}^{2}+\gamma_{2}^{2}} \cos (6\theta -\phi )$ with $\phi = \arctan \left( \gamma_2/\gamma_1\right)$; so, $\theta$ is pinned at $\phi/6+2n\pi/3$ with arbitrary integer $n$. (For a nematic order, $\theta$ and $\theta+\pi$ are equivalent.) In contrast to Cu$_{x}$Bi$_{2}$Se$_{3}$ whose nematic state is possibly  nodal \cite{Fu2014,Yip2013}, Tl$_{2-x}$Mo$_6$Se$_6$ has a nematic state that is fully gapped for any $\theta$. 

Regardless of the nematic angle, a Kramers pair of Majorana flat bands will barbor on the (001) surface, which is guaranteed by a nonzero 1D winding number over $\mathbf{k}_{\perp} = (k_x,k_y)$ \cite{Sato2011,Schnyder2012}. Although the Majorana surface states can be gapped out by disorder which breaks translational symmetry locally, they will be restored after disorder averaging, similar to weak or crystalline topological insulators \cite{Ando2015,Kraus2011}. Interestingly, there exists a kind of disorder which can host Majorana modes locally. At a nematic vortex core, where three degenerate nematic domain walls meet, the Majorana Kramers pair return and pin to it \cite{Wu2017}.

\emph{Discussion.}---The non-symmorphic crystal structure provides a proper electronic base for odd-parity pairing. Under time-reversal and inversion symmetries, the SOC is shown to favor equal-spin pairing and the $E_{2u}$ state in which the triplet $d$ vector is pinned to the $x$-$y$ plane. This 2D representation state would then spontaneously break the rotation symmetry and produce a nematic order, as in Cu$_{x}$Bi$_{2}$Se$_{3}$ \cite{Fu2014,Matano2016,Yonezawa2016} which has being ultimately confirmed by nuclear magnetic resonance (NMR) experiments \cite{Matano2016}. However, NMR measurement would fail to answer the direction of the $d$ vector as in Sr$_2$RuO$_4$ because $E_{2u}$ and $A_{u}$ states are very close in energy ($T_c$ difference being less than 5\%) and external magnetic fields can easily unpin the $d$ vector from the $x$-$y$ plane. Therefore we suggest the proof can be realized in scanning tunneling spectroscopy or phase-sensitive measurements \cite{Nelson2004}. A salient point for this quasi-1D crystal is that it is a type-II superconductor with huge $\kappa$ \cite{Brusetti1988}, thus forbidding vortex formation, so that pure Zeeman effect can be used to study transitions between superconducting states.

 Recently, similar quasi-1D $\mathcal{A}_2$Cr$_3$As$_3$ ($\mathcal{A}$=K, Rb, Cs) superconductors with comparable $T_c$  were reported \cite{Bao2015,Tang2015a,Tang2015b}, and  suggested to be nodal unconventional superconductivity \cite{Tang2015b,Zhi2015,Pang2015,Zhou2017}. Although they share identical crystal structure as $\mathcal{M}_2$Mo$_6$Se$_6$, their different electron valences \cite{Jiang2015} lead to completely different Fermi surface structures, and consequently, distinct superconductivity theories. We also noted an unexpected superconductivity found in Na$_{2-x}$Mo$_6$Se$_6$ in which a large Na deficiency makes the localized system  superconducting \cite{Petrovic2016}.

\begin{acknowledgments}
SMH was supported by the Ministry of Science and Technology (MoST) in Taiwan under Grant No. 105-2112-M-110-014-MY3 and also by the NCTS of Taiwan. He also thanks S.-K. Yip for discussions and helpful suggestions. We would also like to thank Wei-Feng Tsai for a critical reading of the manuscript. The work at Northeastern University was supported by the US Department of Energy (DOE), 
Office of  Science, Basic Energy Sciences grant number DE-FG02-07ER46352 (core research), 
and benefited from  Northeastern University's Advanced Scientific Computation Center (ASCC), 
the NERSC supercomputing  center through DOE grant number DE-AC02-05CH11231, and support 
(applications to layered materials)  from the DOE EFRC: Center for the Computational Design of 
Functional Layered Materials (CCDM) under DE-SC0012575.
H.L. acknowledges the Singapore National Research Foundation for the support under NRF Award No. NRF-NRFF2013-03.

\end{acknowledgments}

\end{document}